\documentclass{jfm}
\usepackage{natbib}
\usepackage{multirow}
\usepackage[table]{xcolor}
\usepackage{rotating}
\usepackage{amsmath}
\usepackage{bm}
\usepackage{array}
\usepackage{graphicx}
\usepackage{dcolumn}   
\usepackage{amssymb}
\usepackage[english]{babel}

\begin{document}
\title{ The boundary integral formulation of Stokes flows includes slender-body theory}
\date{\today}
\author{Lyndon Koens and Eric Lauga\footnote{Email: e.lauga@damtp.cam.ac.uk}}
\affiliation{ Department of Applied Mathematics and Theoretical Physics, University of Cambridge, Wilberforce Road, Cambridge CB3 0WA, United Kingdom}
\maketitle
\begin{abstract} 
The incompressible Stokes equations can classically be recast in a boundary integral (BI) representation, which provides a general method to solve low-Reynolds number problems  analytically and computationally. 
Alternatively, one can solve the Stokes equations by using an appropriate distribution of flow singularities of the right strength within the boundary, a method particularly useful to describe the dynamics of long slender objects for which the numerical implementation of the BI representation becomes cumbersome.  While the BI approach is a mathematical consequence of the Stokes equations, the singularity method involves making judicious guesses that can only be justified a posteriori. In this paper we use matched asymptotic expansions to derive an algebraically accurate slender-body theory directly from the BI representation   able to handle arbitrary surface velocities and surface tractions. This expansion procedure leads to sets of uncoupled linear equations and  to a single one-dimensional integral equation identical to that derived by \citet{Keller1976a} and \citet{Johnson1979} using the singularity method. Hence we show that it is a mathematical consequence of the BI approach that the leading-order flow around a slender body can be represented using a distribution of singularities along its centreline.  Furthermore when derived from either the single-layer or double-layer modified  BI representation, general slender solutions are only possible in certain types of flow, in accordance with the limitations of these representations.

\end{abstract}

\section{Introduction}
The low-Reynolds number hydrodynamics of viscous fluids is accurately captured by the incompressible Stokes equations,
\begin{eqnarray}
-\nabla p +\mu \nabla^{2} \mathbf{u} + \mathbf{f}&=& {\bf 0}\label{Stokes},\\
\nabla \cdot \mathbf{u} &=& 0, \label{incomp}
\end{eqnarray}
where $\mathbf{u}$ is the velocity of the fluid, $p$ is the dynamic pressure, $\mu$ is the dynamic viscosity and $\mathbf{f}$ is any external force density acting on the fluid \citep{Kim2005}.  

The Green's function for these equations, $\mathbi{G}(\mathbf{R})$, is called the stokeslet and represents the flow from a point force of strength $\mathbf{F}$. In an unbounded fluid the associated flow $\mathbf{u}_{S}(\mathbf{x})$ at position $\mathbf{x}$ resulting from a force  located at $\mathbf{x}_{0}$ takes the form
\begin{equation}
8 \pi \mu \mathbf{u}_{S}(\mathbf{x}) = \mathbi{G}(\mathbf{R}) \cdot\mathbf{F}= \frac{\mathbi{I} +\mathbf{\hat{R}}\mathbf{\hat{R}}}{|\mathbf{R}|} \cdot \mathbf{F},
\end{equation}
where  $\mathbi{I}$ is the identity tensor, $\mathbf{R} = \mathbf{x}-\mathbf{x}_{0}$ is a vector from the point force to $\mathbf{x}$, $|\mathbf{R}|$ is the magnitude of $\mathbf{R}$, and $\hat{\mathbf{R}}$ is a unit vector in the direction of $\mathbf{R}$. 
This stokeslet allows the velocity at any point within the fluid, $\mathbf{x}$, to be expressed as \citep{Pozrikidis1992}
\begin{equation}
8 \pi \mu \mathbf{u}(\mathbf{x}) =  \iint_S \,d S(\mathbf{x}_{0}) \left[ \mathbi{G}(\mathbf{R}) \cdot\mathbf{f}(\mathbf{x}_{0})\right]  + \mu \iint_S \,d S(\mathbf{x}_{0}) \left[\mathbf{U}(\mathbf{x}_{0})\cdot \mathbi{T}(\mathbf{R}) \cdot \mathbf{\hat{n}}_{S}(\mathbf{x}_{0}) \right], \label{boundary integral2}
\end{equation}
where  $\mathbf{U}(\mathbf{x}_{0})$ is the surface velocity at $\mathbf{x}_{0}$, $\iint_S \,d S(\mathbf{x}_{0})$ are integrals over the boundaries  $S$ of the fluid, $\mathbf{\hat{n}}_{S}$ is the surface normal pointing out of the fluid, $\mathbf{f}(\mathbf{x}_{0}) = \boldsymbol{\sigma}(\mathbf{x}_{0}) \cdot \mathbf{\hat{n}}_{s}$ is the surface traction on the fluid at $\mathbf{x}_{0}$, $\boldsymbol{\sigma}(\mathbf{x}_{0})$ is the fluid stress at $\mathbf{x}_{0}$ and $\mathbi{T}(\mathbf{R})$ is the   stress tensor generated from the stokeslet \citep{Pozrikidis1992}. In an unbounded fluid the stress tensor, $\mathbi{T}(\mathbf{R})$, has the form
\begin{equation}
 \mathbi{T}(\mathbf{R})= -6 \frac{\mathbf{\hat{R}} \mathbf{\hat{R}} \mathbf{\hat{R}}}{|\mathbf{R}|^{2}}\cdot
 \label{stress}
\end{equation}
In the limit that $\mathbf{x}$ approaches the boundary in Eq.~\eqref{boundary integral2},   the integral equation becomes
\begin{equation}
4 \pi \mu \mathbf{U}(\mathbf{x}) =  \iint_S \,d S(\mathbf{x}_{0}) \left[ \mathbi{G}(\mathbf{R}) \cdot\mathbf{f}(\mathbf{x}_{0})\right]  + \mu \iint_S^{PV} \,d S(\mathbf{x}_{0}) \left[\mathbf{U}(\mathbf{x}_{0})\cdot \mathbi{T}(\mathbf{R}) \cdot \mathbf{\hat{n}}_{S}(\mathbf{x}_{0}) \right], \label{boundary integral}
\end{equation}
where $\iint^{PV} \,d S(\mathbf{x}_{0})$ is the Cauchy principal value of the integral. The above equation provides a relationship between the surface traction, $\mathbf{f}(\mathbf{x}_{0})$, and the surface velocity, $\mathbf{U}(\mathbf{x}_{0})$ and thus, together with Eq.~\eqref{boundary integral2} 
 completely determines the flow. Equations \eqref{boundary integral2} with \eqref{boundary integral} are called the boundary integral (BI) representation of Stokes flow. Note that the first integral on the right-hand side of Eq.~\eqref{boundary integral2} is usually called the single-layer potential while the second integral  is referred to as the double-layer potential.

Alternatively, Stokes flow problems can be solved through the  singularity representation method where fundamental flow singularities are placed outside the fluid region to exactly satisfy all  boundary conditions. For an isolated body in flow, this means that these singularities are placed inside the body. One difficulty of this method is that in general the details of the singularities needed to satisfy the boundary conditions  (type, location and strength) are not known a priori. As a result, exact solutions using this method are only known for simple shapes. For example, \citet{Chwang2006} showed that the rigid body motion of a prolate ellipsoid of arbitrary aspect ratio can be represented by a line distribution of four singularities placed between the foci of the body. In general, however, infinite singularities may be needed to represent the desired flow \citep{Kim2005}.

In the limit of very large aspect ratios, a prolate ellipsoid becomes a long slender body with a straight centreline. For such slender shapes, numerical implementations of the BI integrals tend to require a high resolution to resolve both length scales of the body, and so many singularity representation methods, called slender-body theories (SBTs), have been developed to overcome this difficulty \citep{Cox, Batchelor1970, Clarke1972, 1976, Keller1976a, Johnson1979, Sellier1999, Gotz2000, Koens2016, Koens2017}.  
Early SBTs used a line of stokeslets to represent the rigid body motion of the object, and typically expanded the system in orders of $1/\ln(r_{f}/\ell)$  where $2 \ell$ is the total length of the slender body and $r_{f}$ its maximum radius \citep{Cox, Batchelor1970}.  \citet{Clarke1972} showed that, for straight slender-bodies, the equations of \citet{Batchelor1970} could also be derived by expanding the single-layer BI in powers of $1/\ln(r_{f}/\ell)$.  

Algebraically accurate SBTs were later developed by including higher-order singularities, often inspired by the exact solution for a prolate ellipsoid  \citep{1976, Keller1976a, Johnson1979}. Algebraic SBTs typically represent the flow around a rigid body through a one-dimensional Fredholm equation of the second kind. For example, the SBT by \citet{Keller1976a} can be written as\footnote{Equation~(12) of \citet{Keller1976a}  can be written in the  form of Eq.~\eqref{SBT} by collecting all velocity terms on the left, all force terms on the right, and recognising that $\mathbf{i}= \mathbf{\hat{t}}$, $\boldsymbol{\alpha} = \mathbf{f}$, and $\mathbf{j}$ is a projection operator  taking the components perpendicular to $\mathbf{\hat{t}}$ (i.e.~$\mathbf{j}\mathbf{j} = \mathbi{I} - \mathbf{\hat{t}}\mathbf{\hat{t}}$).} 
\begin{eqnarray}
8 \pi \mathbf{U}(s) &=&   \int_{-\ell}^{\ell}   \left( \frac{\mathbi{I}+ \mathbf{\hat{R}}_{0}\mathbf{\hat{R}}_{0}}{|\mathbf{R}_{0}|} \cdot \mathbf{f}(s') -\frac{\mathbi{I}+ \mathbf{\hat{t}}\mathbf{\hat{t}}}{|s'-s|} \cdot  \mathbf{f}(s) \right)\,d s'  \notag \\
&&  +  \left[L_{SBT} (\mathbi{I}+ \mathbf{\hat{t}}\mathbf{\hat{t}})+ \mathbi{I}- 3 \mathbf{\hat{t}}\mathbf{\hat{t}} \right]\cdot\mathbf{f}(s), \label{SBT}
\end{eqnarray}
where $s\in[-\ell,\ell]$ is the arclength of centreline, $\mathbf{r}(s)$ is the centreline of the slender body, 
$\mathbf{R}_{0}= \mathbf{r}(s)-\mathbf{r}(s')$, $\mathbf{\hat{t}}$ is the unit  vector tangent to the centreline,  $\mathbf{f}(s)$ is the hydrodynamic force density on the filament, $\mathbf{U}(s)$ is the velocity of the body's centreline, 
$L_{SBT} =\ln[4 \ell^{2} (1-s^{2})/(r_{f}^{2}\rho^{2})]$ and 
$r_{f}\rho(s)$ is the radius of the filament at $s$ (Fig.~\ref{fig:slenderbodyconfig}).  
Later, \citet{Johnson1979} derived the same integral equations by placing a distribution of stokeslets and potential source dipoles (defined as half the Laplacian of a stokeslet) along the centreline of the body and expanding the resultant flow in the slender limit. In doing so he also created a set of equations that were accurate to order $(r_{f}/\ell)^{2}$  and showed that $\mathbf{f}(s)$ determined the force per unit length to order $r_{f}^{2} \log(r_{f}/\ell)/ \ell^{2}$. Slender body theories have proved crucial in understanding the hydrodynamics and rheology of filaments \citep{Tornberg2004, Tornberg2006}, the swimming of microorganisms \citep{Barta1988, Myerscough1989, SMITH2009, Koens2014}, and the behaviour of active systems \citep{Guo2018}.

In this paper we demonstrate that a general SBT can be derived directly from the BI representation for Stokes flows  by algebraically expanding the kernels of the BI representation in a matched asymptotic expansion. The equation we obtain relating the mean force per unit length to the centreline velocity is identical to the Keller-Rubinow-Johnson SBT equation (KRJ-SBT), Eq.~\eqref{SBT}, thereby revealing that the BI representation reduces to the singularity representation for slender bodies in the appropriate limits.  These SBT equations are a direct consequence of the Stokes equations, can handle arbitrary surface velocities and forces, and require no assumptions about singularity types, locations or strengths.  The modified single-layer and double-layer BI equations are also addressed, showing that corresponding SBT solutions only exit when specific conditions are met.

 \section{Geometry of slender body}

\begin{figure}
  \centerline{\includegraphics[width=0.9\textwidth]{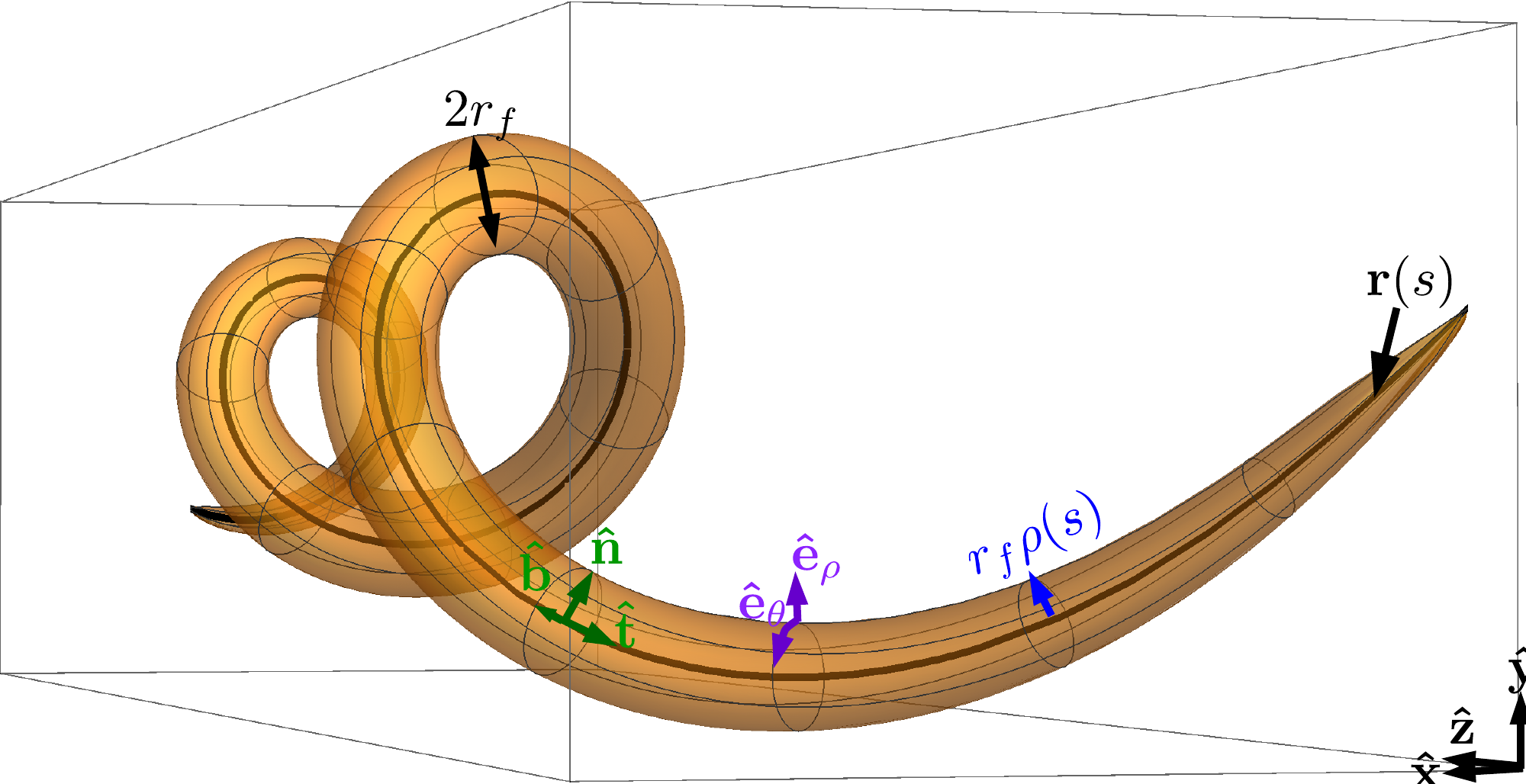}}
  \caption{Geometric representation of a slender body of length $2\ell$ and maximum radius $r_f$. Here $\mathbf{r}(s)$ is the position of the centreline, $r_{f} \rho(s) \in [0,r_{f}]$   the radius of the body, $\mathbf{\hat{t}}$   the tangent to the centreline, $\mathbf{\hat{n}}$ is the normal to the centreline, $\mathbf{\hat{b}}$   the bi-normal to the centreline, $\mathbf{\hat{e}}_{\rho}(s,\theta) =\cos[\theta- \theta_{i}(s)] \mathbf{\hat{n}}(s) + \sin[\theta- \theta_{i}(s)] \mathbf{\hat{b}}(s)$   the radial unit vector and $\mathbf{\hat{e}}_{\theta}(s,\theta) =-\sin[\theta- \theta_{i}(s)] \mathbf{\hat{n}}(s) + \cos[\theta- \theta_{i}(s)] \mathbf{\hat{b}}(s)$   the polar unit vector.}
\label{fig:slenderbodyconfig}
\end{figure}

Points on the surface of a   filamentous body of total arclength $2\ell$ and a circular cross section
of maximum radius $r_{f}$  can be  described as using the arclength $s \in [-\ell, \ell]$ as
\begin{equation}
\mathbf{S}(s,\theta) = \mathbf{r}(s) + r_{f}\rho(s) \mathbf{\hat{e}}_{\rho}(s,\theta), \label{filamentshape}
\end{equation}
where   $\mathbf{r}(s)$ is the centreline of the filament, $r_{f} \rho(s) \in [0,r_{f}]$ is the cross sectional radius of the body at $s$, $\mathbf{\hat{e}}_{\rho}(s,\theta) =\cos[\theta- \theta_{i}(s)] \mathbf{\hat{n}}(s) + \sin[\theta- \theta_{i}(s)] \mathbf{\hat{b}}(s)$ is the local radial vector perpendicular to the centreline tangent $\mathbf{\hat{t}}(s)$, $\mathbf{\hat{n}}(s)$ denotes the normal to the centreline and $\mathbf{\hat{b}}(s)$ is the bi-normal, $\theta \in [-\pi,\pi]$ is the azimuthal angle of the cross section,  and $\theta_{i}(s)$ accounts for the torsion of the curve (Fig.~\ref{fig:slenderbodyconfig}). We choose $\theta_{i}(s)$  to satisfy
\begin{equation}
\frac{d  \theta_{i}}{ds} = \tau(s),
\end{equation}
such that
\begin{equation}
\frac{d \mathbf{\hat{e}}_{\rho}}{d s} = - \kappa(s) \cos[\theta- \theta_{i}(s)] \mathbf{\hat{t}},
\end{equation}
where $\kappa(s)$ and $\tau(s)$ are the curvature and torsion of the centreline respectively.
This parametrisation describes a filament if $r_{f} < \ell$ and a slender body in the limit $r_{f} \ll \ell$.

\section{Asymptotic expansion of the boundary integrals}

In this section we expand the BI representation, Eq.~\eqref{boundary integral}, for  small values of $\epsilon \equiv r_{f}/\ell$ to derive a set of slender-body equations. This is done by performing a matched asymptotic expansion on the integral kernels before evaluating the integrals. A similar derivation was used by \citet{Johnson1979} and \citet{Gotz2000} to derive their respective SBTs but was applied to a prescribed  line of singularities rather than to the full BI equations. As the BI is divided into single-layer and double-layer components, the kernels to expand are
\begin{eqnarray}
\mathbf{K}_{t} &\equiv& \left|\frac{d \mathbf{S}}{d s'}\times\frac{d \mathbf{S}}{d \theta'}\right|\mathbi{G}(\mathbf{R}) \cdot\mathbf{f}(\mathbf{x}_{0}) =  \rho(s') \frac{\mathbi{I}+ \mathbf{\hat{R}}\mathbf{\hat{R}}}{|\mathbf{R}|} \cdot \mathbf{f}(s',\theta'),\label{Kt1}\\
\mathbf{K}_{s} &\equiv& \left|\frac{d \mathbf{S}}{d s'}\times\frac{d \mathbf{S}}{d \theta'}\right|\left[\mathbf{U}(\mathbf{x}_{0})\cdot \mathbi{T}(\mathbf{R}) \cdot \mathbf{\hat{n}}_{s}(\mathbf{x}_{0}) \right]  = - 6\epsilon \rho(s')  \mathbf{U}(s',\theta')\cdot  \frac{\mathbf{\hat{R}} \mathbf{\hat{R}} \mathbf{\hat{R}}}{|\mathbf{R}|^{2}} \cdot \mathbf{\hat{n}}_{S}(s',\theta'),\quad\label{Ks1}
\end{eqnarray}
where $\mathbf{K}_{t}$ is the single-layer kernel, $\mathbf{K}_{s}$ is the double-layer kernel, $\mathbf{U}(s,\theta)$ is the surface velocity of the body at $(s,\theta)$ and
\begin{equation}
\mathbf{R} \equiv \mathbf{S}(s,\theta)-\mathbf{S}(s',\theta') = \mathbf{R}_{0}(s,s') + \epsilon \left[ \rho(s) \mathbf{\hat{e}}_{\rho} (s,\theta) - \rho(s') \mathbf{\hat{e}}_{\rho} (s',\theta')\right].
\end{equation}
 In the above, we have scaled all lengths by half the arc length of the body, $\ell$, velocities by a typical velocity, $U$, and the surface traction by a characteristic force per unit surface area, $\mu \ell U/(r_{f}\ell)$. Furthermore, the slender surface integration factor, 
\begin{eqnarray}
 \left|\frac{\partial \mathbf{S}(s',\theta')}{\partial s'}\times\frac{\partial \mathbf{S}(s',\theta')}{\partial \theta'}\right| = \epsilon \rho(s') + O(\epsilon^{2}),
\end{eqnarray}
has also been included in the kernels in order to reduce the surface integration to a double integral over $s'\in[-1,1]$ and $\theta' \in[-\pi,\pi]$.\footnote{ The integration factor is obtained   by recognising that $\partial_{s} \mathbf{S}(s,\theta) = \mathbf{\hat{t}}(s) + \epsilon \partial_{s} [\rho(s) \mathbf{\hat{e}}_{\rho}(s,\theta)]$, $\partial_{\theta} \mathbf{S}(s,\theta) = \epsilon  \rho(s) \partial_{\theta}\mathbf{\hat{e}}_{\rho}(s,\theta)$, and $\mathbf{\hat{t}} \times \partial_{\theta}\mathbf{\hat{e}}_{\rho} = -\mathbf{\hat{e}}_{\rho}$ and performing a Taylor expansion in $\epsilon$.} Physically the single-layer kernel, $\mathbf{K}_{t}$, describes the flow induced by a distribution of forces located on the surface of the body, while the double-layer kernel, $\mathbf{K}_{s}$, can be interpreted as the flow created by    a distribution of force dipoles and irrotational sources  (see \citealt{Kim2005}).

 In the slender limit, the kernels in Eqs.~\eqref{Kt1}-\eqref{Ks1} have two regions of behaviour: (i) an outer region, where $s-s' = O(1)$, and (ii) an inner region where $s-s' = O(\epsilon)$. We therefore need to expand both kernels in these limits and match their values to determine the leading-order results. Throughout this expansion, higher-order contributions can also be determined by extending the Taylor series expansions further but have been omitted here for simplicity.

\subsection{Outer region}

In the outer region $\mathbf{R}_{0}(s,s')$, is assumed to be of order 1 and so $\mathbf{R} \approx \mathbf{R}_{0}$. Using the superscript $(o)$ to indicate the outer region expansion, the leading-order contribution to kernels in this region are
\begin{eqnarray}
\mathbf{K}_{t} ^{(o)} &=&  \rho(s')\frac{\mathbi{I}+ \mathbf{\hat{R}}_{0}\mathbf{\hat{R}}_{0}}{|\mathbf{R}_{0}|} \cdot \mathbf{f}(s',\theta')+O(\epsilon), \\
\mathbf{K}_{s} ^{(o)} &=& 6\epsilon \rho(s')  \mathbf{U}(s',\theta')\cdot  \frac{\mathbf{\hat{R}}_{0} \mathbf{\hat{R}}_{0} \mathbf{\hat{R}}_{0}}{|\mathbf{R}_{0}|^{2}} \cdot \mathbf{\hat{e}}_{\rho}(s',\theta')+O(\epsilon^{2}), 
\end{eqnarray}
where we have recognised that $\mathbf{\hat{n}}_{S}(s',\theta')=-\mathbf{\hat{e}}_{\rho}(s',\theta')+O(\epsilon)$ for slender bodies. Physically this region captures the influence of surface points \textit{far} from the point of interest, $(s,\theta)$.

\subsection{Inner region} \label{inner}

In the inner region, $s'$ is assumed to be within O($\epsilon$) of $s$. Therefore any dependence on $s'$ can be Taylor expanded around $s$, leading to
\begin{eqnarray}
\mathbf{R}_{0} &=& -\epsilon \chi \mathbf{\hat{t}} + O(\epsilon^{2}), \\
\mathbf{f}(s',\theta') &=& \mathbf{f}(s,\theta')+ O(\epsilon), \\
\rho(s') &=& \rho(s)+ O(\epsilon), \\
\mathbf{\hat{e}}_{\rho}(s',\theta') &=& \mathbf{\hat{e}}_{\rho}(s,\theta')+ O(\epsilon), \\
\mathbf{U}(s',\theta') &=& \mathbf{U}(s,\theta') + O(\epsilon),
\end{eqnarray}
where $\epsilon \chi = (s'-s)$. Using the superscript $(i)$ to indicate  inner-region expansions, the leading-order kernels in the inner region are therefore given by
\begin{eqnarray}
\epsilon \mathbf{K}_{t}^{(i)} &=&  \rho(s) \frac{\mathbi{I}+ \mathbf{\hat{R}}^{(i)}\mathbf{\hat{R}}^{(i)}}{|\mathbf{R}^{(i)}|} \cdot \mathbf{f}(s,\theta')+O(\epsilon), \label{it} \\
\epsilon \mathbf{K}_{s} ^{(i)} &=& 6 \rho(s)  \mathbf{U}(s,\theta')\cdot  \frac{\mathbf{\hat{R}}^{(i)} \mathbf{\hat{R}}^{(i)} \mathbf{\hat{R}}^{(i)}}{|\mathbf{R}^{(i)}|^{2}} \cdot \mathbf{\hat{e}}_{\rho}(s,\theta')+O(\epsilon),  \label{is}
\end{eqnarray}
where
\begin{equation}
\mathbf{R}^{(i)} = - \chi \mathbf{\hat{t}} + \rho(s) [ \mathbf{\hat{e}}_{\rho}(s,\theta)- \mathbf{\hat{e}}_{\rho}(s,\theta')]
= - \chi \mathbf{\hat{t}} + \rho(s) \Delta \mathbf{\hat{e}}_{\rho}(s,\theta,\theta').
\end{equation}
Physically, in this region   the slender body behaves  locally like a cylinder, as reflected in the form of $\mathbf{R}^{(i)}$. As such, the expansion assumes that the curvature, $\kappa(s)$ is much less then $1/\epsilon$, a constraint identical to that assumed by \citet{Johnson1979} and \citet{Gotz2000}. Similar restrictions on the torsion do not occur at this order because of the definition of $\mathbf{\hat{e}}_{\rho}$.

The inner expansion also assumes that the change in the cross section, $\partial_{s} \rho(s)$, is much less than $1/\epsilon$. This condition, however, is likely to be violated near the ends of the body. For example, bodies with ellipsoidal cross sections, $\rho(s) = \sqrt{1-s^{2}}$, have diverging $\partial_{s} \rho(s)$ at their ends. The significance of these end regions to the final integral can estimated using scaling arguments. Consider an end region of  typical size $\sim \sigma$ in which $\epsilon \partial_{s} \rho(s) \sim O(1/\epsilon^{n-1})$ for $n\geq1$. As the leading order inner region is of $O(1/\epsilon)$, the contribution to the BI integral from this region scales as $\sigma \times(1/\epsilon^{n-1}) \times(1/\epsilon) \sim \sigma / \epsilon^{n}$. Hence provided $\sigma / \epsilon^{n} \lesssim O(\epsilon)$ end effects are small. For an ellipsoidal cross section  $\partial_{s} \rho(s)$ becomes of $O(1/\epsilon^{n})$ at $\sigma\sim\epsilon^{2n}$ away from the ends, thereby introducing a correction of $O(\epsilon^{n})$. If, however, $\sigma / \epsilon^{n} \gtrsim O(1)$, a new region in which $\rho(s')$ cannot be expanded around $\rho(s)$ is needed to capture the influence of the ends.

\subsection{Common part}

 Some of the behaviour in the expanded kernels is common to both the outer and inner regions. In order to identify this common behaviour, the relation $\epsilon \chi= s'-s$ is first used to rewrite either the inner region expansion in terms of the outer region variable, $s'$, or the outer region expansion in terms of inner region variable, $\chi$. The common part is then found by expanding the resultant kernel in $\epsilon$. This method, known as the Van Dyke matching method \citep{Hinch1991},   should produce the same result when expanding the outer expansion in the inner variable or vice versa. For our problem, the common part of the BI kernels are
\begin{eqnarray}
\mathbf{K}_{t}^{(i)\in(o)} &=&  \rho(s) \frac{\mathbi{I}+ \mathbf{\hat{t}}\mathbf{\hat{t}}}{|s'-s|} \cdot \mathbf{f}(s,\theta')+O(\epsilon), \\
\mathbf{K}_{s} ^{(i)\in(o)} &=& 6 \epsilon \rho(s')  \mathbf{U}(s,\theta')\cdot  \frac{\mathbf{\hat{t}} \mathbf{\hat{t}} \mathbf{\hat{t}}}{|s'-s|^{2}} \cdot \mathbf{\hat{e}}_{\rho}(s,\theta')+O(\epsilon^{2}), 
\end{eqnarray}
where the superscript $(i)\in(o)$ means we expanded the inner region kernels using the outer region variable (the expansion of the outer region kernel produces the same results).
  
\subsection{Total}

The outer and inner expansions and their common parts can be combined to create a composite representation of each of the BI kernels by adding the outer and inner expansions and then subtracting the common behaviour  \citep{Hinch1991}, thus allowing the composite representation to describe both the leading-order inner and outer  behaviours in a single function. The composite representations of the kernels therefore allow the full BI equations to be approximated by
\begin{equation}
4 \pi  \mathbf{U}(s,\theta) \approx \int_{-1}^{1}  \int_{-\pi}^{\pi}   \left(\mathbf{K}_{t}^{(o)} + \mathbf{K}_{t}^{(i)}-\mathbf{K}_{t}^{(i)\in(o)} +\mathbf{K}_{s}^{(o)} + \mathbf{K}_{s}^{(i)}-\mathbf{K}_{s}^{(i)\in(o)} \right)\,d\theta'\,d s'.
\end{equation}
In this leading-order representation many of the $s'$ integrals take the form
\begin{equation}
\int_{-1}^{1} \frac{\chi^{i}}{\epsilon \sqrt{\chi^{2} + \gamma^{2}}^{j}} \,ds' , \label{inteq}
\end{equation} 
where $\epsilon \chi = s'-s$, $\gamma$ is a constant with respect to $s'$ (can depend on $s$, $\theta$, and $\theta'$) and $i$ and $j$ are positive integers. Integrals of this form  can be evaluated exactly with the substitution $\chi= \gamma \sinh(\phi)$ and then expanded in $\epsilon$ to find the leading-order contribution \citep{Koens2017}. A list of the expansions relevant to this derivation is found in Table~\ref{tab:int}. Significantly, the leading-order component of the asymptotic expansions of Eq.~\eqref{inteq} is also the principle value of these integrals.  Hence the slender-body expansion of the BI representation leads to  
\begin{eqnarray}
4 \pi  \mathbf{U}(s,\theta) &=&  \int_{-1}^{1}   \left( \frac{\mathbf{I}+ \mathbf{\hat{R}}_{0}\mathbf{\hat{R}}_{0}}{|\mathbf{R}_{0}|} \cdot \left\langle \rho(s') \mathbf{f}(s')\right\rangle -\frac{\mathbf{I}+ \mathbf{\hat{t}}\mathbf{\hat{t}}}{|s'-s|} \cdot \left\langle \rho(s)\mathbf{f}(s)\right\rangle \right) \,ds'\notag \\
&& + \int_{-\pi}^{\pi}  \left[L (\mathbf{I}+ \mathbf{\hat{t}}\mathbf{\hat{t}}) - 2 \mathbf{\hat{t}}\mathbf{\hat{t}} + \frac{\Delta \mathbf{\hat{e}}_{\rho} \Delta \mathbf{\hat{e}}_{\rho}}{1-\cos(\theta'-\theta)} \right]\cdot\rho(s)\mathbf{f}(s,\theta')\,d\theta' \notag \\
&& - 2\int_{-\pi}^{\pi}  \left[\mathbf{\hat{t}}\mathbf{\hat{t}} + \frac{\Delta \mathbf{\hat{e}}_{\rho} \Delta \mathbf{\hat{e}}_{\rho}}{1-\cos(\theta'-\theta)} \right] \cdot \mathbf{U}(s,\theta')\,d\theta'+ O(\epsilon), \label{flow}
\end{eqnarray}
where $L= \ln\left(\frac{2 (1-s^{2})}{\epsilon^{2} \rho^{2}(s)[1-\cos(\theta'-\theta)]}\right)$, $\left\langle \mathbf{f}(s')\right\rangle =  \int_{-\pi}^{\pi} \,d\theta' \mathbf{f}(s',\theta')$ and the modulus sign has been dropped from the $1-\cos(\theta'-\theta)$ terms as they are always positive.
  In Eq.~\eqref{flow} the integrals of $\mathbf{K}_{s}^{(o)}-\mathbf{K}_{s}^{(i)\in(o)}$ have been omitted as they are order $\epsilon$ smaller than the other terms. 

\begin{table}
  \begin{center}
  \def~{\hphantom{0}}
\begin{tabular}{l ccc}  
 & $i=0$ & $i=1$ & $i=2$ \\ 
 $j=1$ & $ \log\left( \frac{4(1-s^{2})}{\epsilon^{2} \gamma^{2}}\right) $ & -  &  -   \\ 
$j=3$ & $\frac{2  }{\gamma^{2}} $ & $\frac{2 s \epsilon}{s^{2}-1}$&$ \left[\log\left( \frac{4(1-s^{2})}{\epsilon^{2} \gamma^{2}}\right) -2 \right]$  \\ 
$j=5$ & $\frac{4}{3 \gamma^{4}} $ & $  0$ &$\frac{2 }{3 \gamma^{2}}$ \\
\end{tabular}
\caption{Asymptotic values of the integrals in Eq. \eqref{inteq} for different $i$ and $j$ up to $O(\epsilon^{2})$. 
}
\label{tab:int}
  \end{center}
\end{table}

 Note that the logarithmic term, $L$, diverges at the ends of the body, $|s|\rightarrow 1$, unless $ \rho^{2}(s) \sim (1-s^{2})$. This suggests that ellipsoidal ends are the natural choice for the ends of the bodies in this expansion and indeed for such shapes, 
a Taylor expansion of $\rho^{2}(s)$ near $|s|=1$ leads to $\rho^{2}(s) \rightarrow 0$ and $\partial_{s} \rho^{2}(s) \rightarrow  \partial_{ss} \rho^{2}(s)$. For non-ellipsoidal ends, the equations can still be used as the logarithmic singularity is integrable, however to accurately capture the influence of the ends an improved treatment may be required.
 
\section{The slender-body theory equations}

The slender BI equation, Eq.~\eqref{flow}, can be further simplified by expanding the surface velocity and traction in a Fourier series. Specifically, using the general Fourier expansions
\begin{eqnarray}
 2 \pi \mathbf{U}(s,\theta) &=& \mathbf{U}_{0}(s) + \sum_{n=1}^{\infty} \mathbf{U}_{c,n}(s)\cos[n(\theta-\theta_{i})] +\mathbf{U}_{s,n}(s) \sin[n(\theta-\theta_{i})], \label{Uf}\\
 2 \pi \rho(s)\mathbf{f}(s,\theta) &=& \mathbf{f}_{0}(s) + \sum_{n=1}^{\infty} \mathbf{f}_{c,n}(s)\cos[n(\theta-\theta_{i})] +\mathbf{f}_{s,n}(s) \sin[n(\theta-\theta_{i})], \label{Ff}
\end{eqnarray}
combined with the integral identities \citep{I.S.GradshteynAuthorI.M.RyzhikAuthorAlanJeffreyAuthor2000}
\begin{eqnarray}
\int_{0}^{2\pi} \,d \theta \ln\left[ 1-\cos(\theta)\right] &=& - 2\pi \log(2),  \\
\int_{0}^{2\pi} \,d \theta \ln\left[ 1-\cos(\theta)\right] \cos(n \theta) &=& -\frac{2 \pi}{n},\quad (n> 0), \\
\int_{0}^{2\pi} \,d \theta \ln\left[ 1-\cos(\theta)\right] \sin(n \theta) &=& 0,\quad (n\geq 0),
\end{eqnarray}
allows all the integrals involving $\theta'$ to be evaluated. For these Fourier expansions, subscript $0$ represents the zeroth mode, subscript $c,n$ represents the $n$th cosine mode and subscript $s,n$ represents the $n$th sine mode. The orthogonality of trigonometric functions then reduces the zeroth-order Fourier mode of Eq.~\eqref{flow} to
\begin{eqnarray}
4 \mathbf{U}_{0}(s) &=&   \int_{-1}^{1} \,d s'  \left( \frac{\mathbf{I}+ \mathbf{\hat{R}}_{0}\mathbf{\hat{R}}_{0}}{|\mathbf{R}_{0}|} \cdot \mathbf{f}_{0}(s') -\frac{\mathbf{I}+ \mathbf{\hat{t}}\mathbf{\hat{t}}}{|s'-s|} \cdot  \mathbf{f}_{0}(s) \right)  \notag \\
&&  +  \left[L_{SBT} (\mathbf{I}+ \mathbf{\hat{t}}\mathbf{\hat{t}})+ \mathbf{I}- 3 \mathbf{\hat{t}}\mathbf{\hat{t}} \right]\cdot\mathbf{f}_{0}(s) +O(\epsilon), \label{JSBT}
\end{eqnarray}
while the higher-order Fourier modes lead to
\begin{eqnarray}
2 n \mathbf{U}_{c,n}(s) &=& (\mathbf{I}+ \mathbf{\hat{t}}\mathbf{\hat{t}}) \cdot \mathbf{f}_{c,n}(s) +O(\epsilon),\quad (n> 0) , \label{CSBT} \\
2 n \mathbf{U}_{s,n}(s) &=& (\mathbf{I}+ \mathbf{\hat{t}}\mathbf{\hat{t}}) \cdot \mathbf{f}_{s,n}(s) +O(\epsilon),\quad (n> 0), \label{SSBT}
\end{eqnarray}
where $L_{SBT} =\ln\left[4 (1-s^{2})/(\epsilon^{2} \rho^{2}(s))\right]$.
The equation  above provides a relationship between the movement of the centreline and the average force per unit length on the fluid from the body, while the second and third relate more complex surface deformations to the corresponding surface tractions.  Similarly to other slender-body theory treatments, these relationships are either linear or a one-dimensional Fredholm equation of the second kind. Solutions to a boundary value problem are therefore determined by expanding the known surface distribution in a Fourier series in $\theta$, inserting the Fourier coefficients into Eqs.~\eqref{JSBT}-\eqref{SSBT} to determine the coefficients of the unknown distribution and then reconstructing that distribution from Eqs.~\eqref{Uf} and \eqref{Ff}.
 Importantly, the equation for the mean force per unit length, Eq.~\eqref{JSBT}, is identical to the the KRJ-SBT equation, Eq.~\eqref{SBT}. \citet{Johnson1979} originally derived this equation by expanding the flow from a line distribution of stokeslets and potential source dipoles, and so  this singularity representation captures the same physics as the BI representation when both are expanded in the slender body limit.

A similar equivalence may be found for the torque per unit length generated from the surface rotation. For a cylindrical body, the surface velocity, surface traction, and torque per unit length, $\mathbf{l}(s)$, resulting from the local rotation $\Omega \mathbf{\hat{t}}$ are given by
\begin{eqnarray}
\mathbf{U}(s,\theta) &=& \epsilon \rho(s)\Omega \mathbf{\hat{t}}\times\mathbf{\mathbf{\hat{e}}}_{\rho}(s,\theta) = \epsilon \rho(s)\Omega\mathbf{\mathbf{\hat{e}}}_{\theta}(s,\theta), \\
\rho(s)\mathbf{f}(s,\theta) &=& 2\epsilon \rho(s) \Omega \mathbf{\mathbf{\hat{e}}}_{\theta}(s,\theta), \\
\mathbf{l}(s) &=& \int_{-\pi}^{\pi} \,d\theta \rho(s) \left[\mathbf{S}(s,\theta)\times \mathbf{f}(s,\theta)\right] = 4 \pi \epsilon^{2} \rho^{2}(s) \Omega \mathbf{\hat{t}}, \label{torque}
\end{eqnarray}
where $ \mathbf{\mathbf{\hat{e}}}_{\theta}(s,\theta)=- \sin[\theta- \theta_{i}(s)] \mathbf{\hat{n}}+ \cos[\theta- \theta_{i}(s)] \mathbf{\hat{b}}$.
The above torque is identical to that derived from a line of singularities \citep{Koens2014}, showing again that the singularity and boundary integral representations   capture the same physics in the slender-body limit.

\section{Single-layer and double-layer slender-body equations}

Under certain  conditions, the single-layer and double-layer potentials are also separetely solutions to the Stokes equations and may be used to form modified BI equations \citep{Pozrikidis1992}. Specifically, the single-layer BI representation is written as
\begin{equation}
8 \pi \mu \mathbf{U}(\mathbf{x}) = \iint_S \,d S(\mathbf{x}_{0}) \left[ \mathbf{G}(\mathbf{R}) \cdot\mathbf{f}^{m}(\mathbf{x}_{0})\right], \label{1layerintegral}
\end{equation}
and holds when there is no flux of fluid through the boundaries.  The double-layer BI is
\begin{equation}
 8 \pi \mathbf{U}(\mathbf{x}) = 4 \pi \mathbf{U}^{m}(\mathbf{x})+ \iint_S^{PV} \,d S(\mathbf{x}_{0}) \left[\mathbf{U}^{m}(\mathbf{x}_{0})\cdot \mathbf{T}(\mathbf{R}) \cdot \mathbf{\hat{n}}_{s}(\mathbf{x}_{0}) \right], \label{2layerintegral}
\end{equation}
and is valid when there is no net force or torque on bodies within the boundaries. In these equations, $\mathbf{f}^{m}$ is termed the modified surface traction on the fluid and $\mathbf{U}^{m}$ is the modified surface velocity. In numerical computations, these equations are often easier to evaluate then the full BI. It is therefore relevant to consider what slender-body theory equations would arise from these two representations.

The general slender BI equations, Eq.~\eqref{flow}, contains the expanded forms of the single-layer and double-layer potentials. 
The single-layer representation can, therefore, be found by replacing the $4\pi$ on the left-hand side with $8\pi$, the surface traction with the modified traction, $\mathbf{f} \rightarrow \mathbf{f}^{m}$, and the velocities on the right-hand side with zero, $\mathbf{U} = 0$. The resultant single-layer SBT equations is given for the mean as
\begin{eqnarray}
4 \mathbf{U}_{0}(s) &=&   \int_{-1}^{1} \,d s'  \left( \frac{\mathbf{I}+ \mathbf{\hat{R}}_{0}\mathbf{\hat{R}}_{0}}{|\mathbf{R}_{0}|} \cdot \mathbf{f}^{m}_{0}(s') -\frac{\mathbf{I}+ \mathbf{\hat{t}}\mathbf{\hat{t}}}{|s'-s|} \cdot  \mathbf{f}^{m}_{0}(s) \right)  \notag \\
&&  +  \left[L_{SBT} (\mathbf{I}+ \mathbf{\hat{t}}\mathbf{\hat{t}})+ \mathbf{I}- 3 \mathbf{\hat{t}}\mathbf{\hat{t}} \right]\cdot\mathbf{f}^{m}_{0}(s) +O(\epsilon),
\end{eqnarray}
while for the other Fourier modes we now have
\begin{eqnarray}
8  \mathbf{U}_{c,1}(s) &=& \left[2(\mathbf{I}+ \mathbf{\hat{t}}\mathbf{\hat{t}}) - \mathbf{\hat{n}}\mathbf{\hat{n}} + \mathbf{\hat{b}}\mathbf{\hat{b}}\right]\cdot \mathbf{f}^{m}_{c,1}(s) - \left(  \mathbf{\hat{n}}\mathbf{\hat{b}} +\mathbf{\hat{b}}\mathbf{\hat{n}}\right)\cdot \mathbf{f}^{m}_{s,1}(s)+O(\epsilon), \\
8  \mathbf{U}_{s,1}(s) &=& \left[2(\mathbf{I}+ \mathbf{\hat{t}}\mathbf{\hat{t}}) + \mathbf{\hat{n}}\mathbf{\hat{n}} - \mathbf{\hat{b}}\mathbf{\hat{b}}\right]\cdot \mathbf{f}^{m}_{s,1}(s) - \left(  \mathbf{\hat{n}}\mathbf{\hat{b}} +\mathbf{\hat{b}}\mathbf{\hat{n}}\right)\cdot \mathbf{f}^{m}_{c,1}(s)+O(\epsilon), \\
4 n \mathbf{U}_{c,n}(s) &=& (\mathbf{I}+ \mathbf{\hat{t}}\mathbf{\hat{t}}) \cdot \mathbf{f}^{m}_{c,n}(s) +O(\epsilon),\quad (n> 1), \\
4 n \mathbf{U}_{s,n}(s) &=& (\mathbf{I}+ \mathbf{\hat{t}}\mathbf{\hat{t}}) \cdot \mathbf{f}^{m}_{s,n}(s) +O(\epsilon) 
,\quad (n> 1),\end{eqnarray}
where  we have   written the modified traction and the surface velocity as Fourier series. 

The equation for the mean force per unit length  is structured exactly like KRJ-SBT; the torque per unit length from surface rotation can be shown to also be the same as Eq.~\eqref{torque}. The difference between the single-layer and full BI representations occurs for the $n=1$ Fourier terms. In the full BI, every force mode was uniquely related to a surface deformation. However in the the single-layer representation, the $n=1$ equations do not fully determine $\mathbf{f}^{m}_{c,1}$ and $\mathbf{f}^{m}_{s,1}$ since these equations have no solution for $\mathbf{f}^{m}_{c,1}\cdot \mathbf{\hat{n}}$ and $\mathbf{f}^{m}_{s,1}\cdot \mathbf{\hat{b}}$ unless $\mathbf{U}_{c,1}\cdot \mathbf{\hat{n}} + \mathbf{U}_{s,1}\cdot \mathbf{\hat{b}}= 0$. This restriction implies that solutions exist only if $\int_{-\pi}^{\pi} \,d \theta \mathbf{U}\cdot\mathbf{\hat{e}}_{\rho} = 0 $, thereby preventing any local changes in  the body volume. The inability of the single-layer SBT to uniquely determine the $n=1$ modes reflects the  inability of the single-layer BI to describe bodies with changing volume \citep{Pozrikidis1992}.
 
Similarly, the double-layer SBT equations can be obtained from Eq.~\eqref{flow} by setting the traction to zero, $\mathbf{f} = \mathbf{0}$, replacing the velocity on the right-hand side of equations with the modified surface velocity $\mathbf{U} \rightarrow \mathbf{U}^{m}$, adding $4 \pi\mathbf{U}^{m}$ to the right-hand side and replacing the $4\pi$ with $8 \pi$ on the left-hand side. The double-layer SBT equations then become
\begin{eqnarray}
4 \mathbf{U}_{0}(s) &=&  \mathbf{0} +O(\epsilon), \\
4  \mathbf{U}_{c,1}(s) &=& \left[2 \mathbf{I} +\mathbf{\hat{n}}\mathbf{\hat{n}} - \mathbf{\hat{b}}\mathbf{\hat{b}}\right]\cdot \mathbf{U}^{m}_{c,1}(s) + \left(  \mathbf{\hat{n}}\mathbf{\hat{b}} +\mathbf{\hat{b}}\mathbf{\hat{n}}\right)\cdot \mathbf{U}^{m}_{s,1}(s)+O(\epsilon), \\
4  \mathbf{U}_{s,1}(s) &=& \left[2 \mathbf{I} + \mathbf{\hat{b}}\mathbf{\hat{b}} -\mathbf{\hat{n}}\mathbf{\hat{n}}\right]\cdot \mathbf{U}^{m}_{s,1}(s) + \left(  \mathbf{\hat{n}}\mathbf{\hat{b}} +\mathbf{\hat{b}}\mathbf{\hat{n}}\right)\cdot \mathbf{U}^{m}_{c,1}(s)+O(\epsilon), \\
4 \mathbf{U}_{c,n}(s) &=& 2  \mathbf{U}^{m}_{c,n}+O(\epsilon),\quad (n> 1), \\
4 \mathbf{U}_{s,n}(s) &=&  2  \mathbf{U}^{m}_{s,n} +O(\epsilon),\quad (n> 1),
\end{eqnarray}
where we have again expanded the functions in Fourier components. Similarly to the single-layer slender-body equations, the double-layer equations  have solutions only if $\mathbf{U}_{0} = \mathbf{0}$ and $\int_{-\pi}^{\pi} \,d \theta \mathbf{U}\cdot\mathbf{\hat{e}}_{\theta} = 0 $, reflecting the inability of the double-layer BI to describe bodies with either net forces or torques.

\section{Conclusion}

The Green's function for the Stokes equations allows low-Reynolds flows to be represented mathematically using either boundary integrals or distributions of singularities. Boundary integrals are an effective general procedure for many shapes but are difficult to implement computationally for slender bodies. In contrast,  singularity representations require  suitable guesses of the relevant singularities but are useful in order to construct theories for the hydrodynamics of long slender bodies (denoted slender-body theories).  

In this paper  we derive an algebraically-accurate slender-body theory directly from the boundary integral representation for Stokes flows. This derivation (i) requires no consideration about the type, strength or location of the singularities, (ii) is a direct result of the Stokes equations and (iii) provides a slender-body theory that relates arbitrary surface velocities to arbitrary surface tractions. The equation for the mean force and torque per unit length has the same structure as the slender-body theories of \citet{Keller1976a} and \citet{Johnson1979}, demonstrating that a representation consisting of placing singularities over the centreline of a body and the boundary integral representation  describe the same physics when expanded in the slender-body limit. 

Our derivation was first obtained from the most general boundary integral representation before being adapted to bother the single-layer and double-layer boundary representations. The equations for the mean force per unit length and mean torque per unit length are the same across the different representations. While the full boundary integral is able to handle any surface velocity, it was found that the single-layer representation only has solutions if the body does not locally change volume and the double-layer representation only has solutions if there is no local translation or surface rotation of the filament. Both of these conditions are the slender-body theory equivalents of the restrictions on modified boundary integrals.  Importantly, as this theory handles arbitrary surface velocities and tractions, this formalism can be applied to non-rigid bodies and bodies near neighbours by including the relevant additional terms.\\

This project has received funding from the European Research Council (ERC) under the European Union's Horizon 2020 research and innovation programme  (grant agreement 682754 to EL).

\bibliographystyle{jfm}
\bibliography{library}

\end{document}